\documentclass[12pt]{article}
\usepackage{amsmath,amsthm,amssymb,euscript,epsfig,youngtab}
\usepackage{array}
\usepackage{fancybox}
\usepackage[usenames]{color}
\usepackage{amsfonts,bm}
\usepackage{graphicx}

\newcommand{\be}{\begin{equation}}
\newcommand{\ee}{\end{equation}}
\newcommand{\bea}{\begin{eqnarray}\displaystyle}
\newcommand{\eea}{\end{eqnarray}}

\makeatletter
\@addtoreset{equation}{section}
\makeatother
\renewcommand{\theequation}{\thesection.\arabic{equation}}
\def\one{{\hbox{ 1\kern-.8mm l}}}
\def\zero{{\hbox{ 0\kern-1.5mm 0}}}

\begin{document}

\makeatletter
\@addtoreset{equation}{section}
\makeatother
\renewcommand{\theequation}{\thesection.\arabic{equation}}

\rightline{WITS-CTP-082}
\vspace{1.8truecm}

\vspace{15pt}

%%%%%%%%%%%%%%%%%

{\LARGE
\centerline{\bf  From Large $N$ Nonplanar Anomalous Dimensions }
\centerline{\bf to Open Spring Theory} 
}  

\vskip.5cm 

\thispagestyle{empty} \centerline{
    {\large \bf Robert de Mello Koch\footnote{{\tt robert@neo.phys.wits.ac.za}},}
   {\large \bf Garreth Kemp\footnote{{\tt Garreth.Kemp@students.wits.ac.za}} and Stephanie Smith\footnote{{\tt Stephanie.Smith@students.wits.ac.za}}}
                                                       }

\vspace{.4cm}
\centerline{{\it National Institute for Theoretical Physics ,}}
\centerline{{\it Department of Physics and Centre for Theoretical Physics }}
\centerline{{\it University of Witwatersrand, Wits, 2050, } }
\centerline{{\it South Africa } }

\vspace{1.4truecm}

%%%%%%%%%%%%%%%%%
\thispagestyle{empty}

\centerline{\bf ABSTRACT}

\vskip.4cm 

In this note we compute the non-planar one loop anomalous dimension of restricted Schur polynomials that
have a bare dimension of $O(N)$. This is achieved by mapping the restricted Schur polynomials into states
of a specific $U(N)$ irreducible representation. In this way the dilatation operator is mapped into a $u(n)$
valued operator and, as a result, can easily be diagonalized. The resulting spectrum is reproduced by a classical
model of springs between masses.

\setcounter{page}{0}
\setcounter{tocdepth}{2}

\newpage

\setcounter{footnote}{0}

\linespread{1.1}
\parskip 4pt

\section{Introduction}

According to the AdS/CFT correspondence\cite{Maldacena:1997re}, the conformal dimension of an operator 
in the ${\cal N}=4$ super Yang-Mills theory maps into the energy of the corresponding state in IIB string 
theory on the AdS$_5\times$S$^5$ background. In this article we are interested in computing the energies
of excited giant graviton systems in string theory by computing the anomalous dimensions of restricted 
Schur polynomials\cite{Balasubramanian:2004nb,Bhattacharyya:2008rb,Bhattacharyya:2008xy} in ${\cal N}=4$
super Yang-Mills theory. These operators have a
classical dimension which is order $N$ and consequently summing the planar diagrams does not give an
accurate large $N$ approximation\cite{Balasubramanian:2001nh}. Fortunately, in the last few years,
starting from \cite{Corley:2001zk,Corley:2002mj}
methods to study the large $N$ limit of such correlators have been 
developed\cite{Bhattacharyya:2008rb,Bhattacharyya:2008xy,Kimura:2007wy,Brown:2007xh,Brown:2008ij,Kimura:2008ac}.
In particular, there are now powerful methods\cite{Koch:2010gp,DeComarmond:2010ie,Carlson:2011hy,Koch:2011jk,Koch:2011hb}
to evaluate the action of the one loop dilatation operator in the $su(2)$
sector\cite{Beisert:2003tq}.

\begin{figure}[ht]%
\begin{center}
\includegraphics[width=9.0cm]{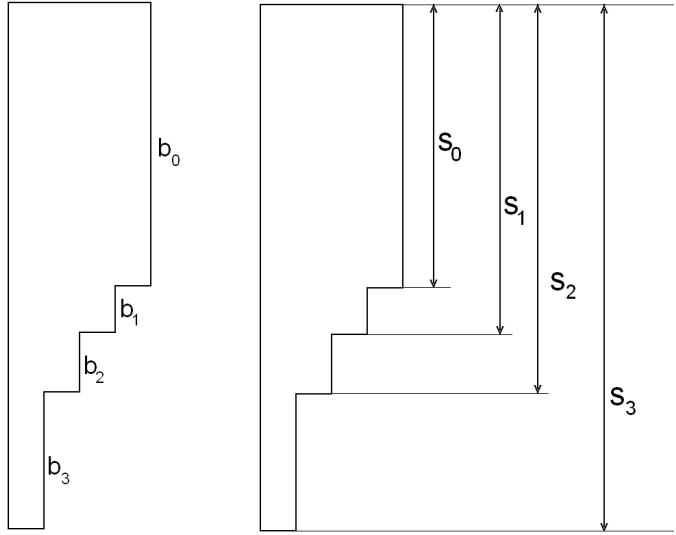}%
\caption{Definition of the $b_i$s and $s_i$s in terms of a Young diagram for $c=4$ columns. 
The relation between the $s_i$ and the $b_i$ is easily read from the figure. For example, $s_2=b_0+b_1+b_2$.
Columns are ordered so that column length increases. They are then numbered starting from 0. 
For the Young diagram shown, the right most column is column 0 and the left most is column 3.
The generalization to any $c$ should be obvious.}%
\label{fig:youngdiag}%
\end{center}
\end{figure}

In this note we will consider the diagonalization of the one loop dilatation operator when acting on restricted
Schur polynomials $\chi_{R,(r,s)}(Z,Y)$ built from $n$ $Z$ fields and $m$ $Y$ fields, with $m\ll n$ and
$m,n$ both order $N$, as in \cite{Koch:2011hb}.
For a system of $p$ sphere giant gravitons, $R$ is a Young diagram with $p$ columns and $m+n$ boxes, $r$
is a Young diagram with $p$ columns and $n$ boxes and $s$ is a Young diagram with at most $p$ columns. 
After diagonalizing on the $s$ label, \cite{Koch:2011hb} finds that the resulting equations for the action
of the dilatation operator can be labeled by configurations of open strings that are consistent with the Gauss 
Law, as well as labels specifying Young diagram $r$, as defined in Figure \ref{fig:youngdiag}. 
For the configuration $C$ with $n_{ij}$ open strings stretching between branes $i$ and $j$ the one loop
dilatation operator is given by
\bea
   DO_C(\{s_i\})=-g_{YM}^2\sum_{\alpha\beta}\, n_{\alpha\beta}\Delta_{\alpha\beta}\, O_C(\{s_i\} )
\label{todiagonalize}
\eea
where the operator $\Delta_{ij}$ acts as follows ($\Delta_{ij}$ only changes the values of
$s_i$ and $s_j$ so that these are the only two variables that we display in the next equation)
$$
  \Delta_{ij}O_C(s_i,s_j) = -(c_i+c_j)O_C(s_i,s_j)+\sqrt{c_ic_j}(O_C(s_i+1,s_j-1)+O_C(s_i-1,s_j+1))\, .
$$
In this last equation $c_a$ is the factor of the last box in column $a$. Recall that a box in row $i$ and
column $j$ has a factor $N-i+j$. The primary goal of this article is to explain how to
diagonalize (\ref{todiagonalize}). This is achieved by mapping the operators $O_C(s_i,s_j)$ into states in the
carrier space of a specific $U(N)$ irreducible representation. The dilatation operator is mapped into a $u(n)$
valued operator and, as a result, can easily be diagonalized. We then go on to show that the resulting spectrum 
is reproduced by a classical model of springs between masses. 

\section{Nonplanar Dilatation Operator}

To start we will review a few elementary facts, familiar from angular momentum
in quantum mechanics, that will play an important role later.
The fundamental representation of $u(N)$ represents the elements of the
Lie algebra as $N\times N$ matrices. The generators can be taken as
$$
  (E_{kl})_{ab}=\delta_{ak}\delta_{bl},\qquad k,l,a,b=1,2,...,N\, .
$$
We will study the operators (the labeling is such that $i>j$ i.e. $Q_{ij}$ is not defined if $i<j$)
$$
  Q_{ij}={E_{ii}-E_{jj}\over 2},\qquad Q^+_{ij}=E_{ij},\qquad Q^-_{ij}=E_{ji}\, ,
$$
which obey the familiar algebra of angular momentum raising and lowering operators
$$
  \big[ Q_{ij},Q_{ij}^+\big]   = Q^+_{ij},\qquad
  \big[ Q_{ij},Q_{ij}^-\big]   =-Q^-_{ij},\qquad
  \big[ Q^+_{ij},Q_{ij}^-\big] = 2Q_{ij}\, .
$$
Although these commutators have been computed making use of the fundamental
representation, we know that they would be the same if they had been computed
in any representation and they define the representation independent Lie algebra.

General representations of these $su(2)$ subalgebras can be labeled with the eigenvalue of
$$
  L_{ij}^2\equiv Q_{ij}^-Q_{ij}^++Q_{ij}^2+Q_{ij}=
          Q_{ij}^+Q_{ij}^-+Q_{ij}^2-Q_{ij}
$$
and states in the representation are labeled by the eigenvalue of $Q_{ij}$
$$
  Q_{ij}|\lambda,\Lambda\rangle = \lambda |\lambda, \Lambda\rangle\, ,
  \qquad
  L_{ij}^2|\lambda,\Lambda\rangle = (\Lambda^2 + \Lambda) |\lambda, \Lambda\rangle\, ,
  \qquad
 -\Lambda\le\lambda\le\Lambda  \, .
$$
Recall that
$$
  Q_{ij}^+ |\lambda,\Lambda\rangle = c_+ |\lambda + 1,\Lambda\rangle\, ,
  \qquad
  c_+=\sqrt{(\Lambda +\lambda +1)(\Lambda-\lambda)}\, , $$
and 
$$
  Q_{ij}^- |\lambda,\Lambda\rangle = c_- |\lambda - 1,\Lambda\rangle\, ,
  \qquad
  c_-=\sqrt{(\Lambda +\lambda)(\Lambda-\lambda+1)} \, .$$

The $N$ operators $E_{ii}$ commute so that we can always choose a basis in which they are simultaneously diagonal. 
Recall the definition of $b_i$ $i=0,1,...,c-1$ for a Young diagram with $c$ columns, given in Figure \ref{fig:youngdiag}.
The restricted Schur polynomials labeled by the Young diagram shown is identified with the state with 
$E_{ii}= 2 (\, N-s_i \, )\, .$ The advantage of identifying the restricted Schur polynomials
with states of a $U(N)$ representation is that we can now write the dilatation operator as a $u(N)$ valued
operator. In particular, the operators $\Delta_{ij}$ are
$$
  \Delta_{ij}=-{1\over 2}(E_{ii}+E_{jj})+Q_{ij}^- +Q_{ij}^+\, .
$$
For simplicity we will now focus on the case $c=2$. In this case, identify
$$
  c_- =\sqrt{(N-b_0)(N-b_0-b_1+1)},\qquad c_+=\sqrt{(N-b_0+1)(N-b_0-b_1)}
$$
so that
$$
  \Lambda={1\over 2} b_{1,{\rm max}},\qquad \lambda ={1\over 2} b_1\, .
$$
We will focus on $b_{1,{\rm max}}$ even so that $\Lambda$ is integer. 
Not all states of the irreducible representation participate: because $b_1\ge 0$ we have $\lambda\ge 0$.
Thus, of the $2b_{1,{\rm max}}+1$ states, only $b_{1,{\rm max}}+1$ of them remain.
Finally, we are interested in the limit $b_{1,{\rm max}}\sim\sqrt{N}$ with $N\to\infty$.  
It is only in this limit that (\ref{todiagonalize}) holds. Away from this limit (\ref{todiagonalize}) picks
up corrections of order $1/b_{1,{\rm max}}$\cite{Koch:2011hb}. There is an obvious extension of this discussion
for $c>2$.

\section{Strings between 2 giants} 

\begin{figure}[ht]
\begin{center}
\includegraphics[width=6.0cm]{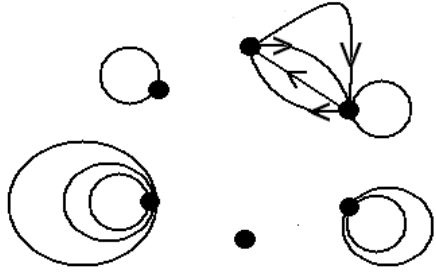}
\caption{The label $C$ for a system of 6 giants. $2n_{12}=4$ strings stretch between branes 1 and 2. There is one
more string attached to brane 2. Two strings are attached to brane 3, 3 strings to brane 5 and a single string to 
brane 6. The dilatation operator action depends only on the strings stretching between different branes\cite{Koch:2011hb}.}
\label{fig:giantsys}
\end{center}
\end{figure}

Consider a system of $p$-giants with $p$ arbitrary except that we fix it to be $O(1)$. The Young diagrams relevant
for these states have $p$ columns. Consider the situation
for which we have $2 n_{ij}$ strings stretching between giants $i$ and $j$. See Figure \ref{fig:giantsys} for an
example of the label $C$ when $p=6$ and $2 n_{ij}=4$. The results of this section are also directly applicable to
the case that pairs of mutually distinct branes have strings stretching between them. In this case, the action
of the dilatation operator is given by a sum of terms which commute and can each be diagonalized using the same method.

\noindent
{\bf Construction of Creation and Annihilation Operators:}
In this case
$$
  D = -2 n_{ij} g_{YM}^2 \Delta_{ij}\, .
$$
For a creation operator we want
\bea
\label{CO}
\big[ D,A^\dagger\big] = \alpha A^\dagger
\eea
with $\alpha > 0$. Make the ansatz
$$
  A^\dagger = aE_{ii}+bE_{jj}+cE_{ij}+dE_{ji}\, .
$$
It is straight forward to verify that (\ref{CO}) implies
$$
  A^\dagger = {1\over 2} (E_{ii} -E_{jj})+{1\over 2} E_{ij}-{1\over 2} E_{ji}
$$
and $\alpha = 4 n_{ij} g_{YM}^2$. To implement the condition $b_1 >0$ we need to require that the oscillator wave function 
has a node at the origin - thus only odd parity (i.e. odd under $b_1\to -b_1$) states are kept. This implies that
half the states are kept so that we land up with a frequency of $8n_{ij}g_{YM}^2$. For $n_{ij}=1$ this is in 
complete agreement with spectrum computed in \cite{Koch:2010gp,DeComarmond:2010ie}. Thus, the spectrum of the dilatation operator is
$$
  \lambda = (8n_{ij}g_{YM}^2)n
$$
with $n$ a not negative integer. This is in complete agreement with the spectrum computed in \cite{Carlson:2011hy}.
There is a simple algebra obeyed by the creation and annihilation operators of this oscillator
$$
  \big[ A,A^\dagger\big]= {1\over 2} (E_{ii}+E_{jj}) + \Delta_{ij} =2N-2b_0-b_1-{D\over 2g_{YM}^2}
                        = b_{1,{\rm max}}-{D\over 2g_{YM}^2}\, .
$$
If we introduce the oscillators $A=\sqrt{b_{1,{\rm max}}} a$
we find, for any state of finite energy in the $b_{1,{\rm max}}\to\infty$ limit
$$
  \big[ a,a^\dagger\big]= 1-{D\over 2 b_{1,{\rm max}} g_{YM}^2}=1\, .
$$

\noindent
{\bf Connection to Continuum Limit:}
We can ask how this compares to the frequencies computed after we have taken the continuum limit of the $\Delta_{ij}$,
described in appendix H of \cite{Koch:2011hb}. From that appendix, we find
$$
 D=-2g_{YM}^2 n_{ij}M_{ab}\left({\partial\over\partial x_a}{\partial\over\partial x_b}-{x_a x_b\over 4}\right)
$$
with
$$
M_{11}=M_{22}=1,\qquad M_{ij}=M_{ji}=-1.
%M=
%\left[
%\begin{array}{cc}
%1 &-1\\
%-1 &1 
%\end{array}
%\right]
$$
The two frequencies are $4n_{ij}g_{YM}^2$ and $0$. The zero frequency corresponds to the motion of the center of mass
($x_{cm}\propto x_i+x_j$). Fix this center of mass motion because the system of giants is fixed. The nonzero frequency 
reproduces what we found above, again after dropping half the states. Clearly then, the continuum limit catches the 
complete large $b_{1,{\rm max}}$ dynamics.

\noindent
{\bf Classical Model:} 
The operators we study are nearly supersymmetric so that it is natural to expect that they correspond to fast moving
strings on the D-brane. It is thus natural to associate them with null trajectories in AdS$_5\times$S$^5$ that are
contained in the D-brane worldvolume. This analysis has been performed in \cite{Berenstein:2006qk}. See
\cite{Berenstein:2005fa,Correa:2006yu} for additional relevant and useful discussion.
The resulting null trajectory leads to a pp-wave and the light cone Hamiltonian is related to the anomalous 
dimension
$$
  H_{\rm light\, cone}={1\over P^+}H_{\perp}=\Delta -n_Z-n_Y = D
$$
where $H_{\perp}$ describes string oscillations in the perpendicular (to string motion) directions and 
$n_Z (n_Y)$ are the number of $Z$s ($Y$s) in the operator.
See also \cite{KS} which is relevant to our discussion. What should we use for $H_{\perp}$?
When we change the number of $Z$'s in the giant we change the radius of the circle on which it is orbiting; this
corresponds to the direction transverse to the giants direction of motion - i.e. the oscillator that we have
diagonalized above is describing oscillations in the perpendicular (to string motion) directions.
The Gauss Law picture of \cite{Koch:2011hb} suggests that the configuration we study consists
of $2n_{12}$ strings stretching between the two giants. Each string is a single $Y$ - so these are short strings that
we will model as two endpoints. The spring constant for springs connected in parallel is the sum of the individual 
spring constants. Thus, the configuration we study will have $k\propto n_{12}$. The scale of the anomalous dimension is
set by $g_{YM}^2$. Under AdS/CFT the anomalous dimension maps to an energy, so that $g_{YM}^2$ naturally sets
the energy scale. To ensure that the scale of the potential energy is set by $g_{YM}^2$ we will choose the
spring constant $k\propto g_{YM}^2$. Making a choice of a constant that will prove to be convenient below,
we set $k=4g_{YM}^2 n_{12}$. Adding a kinetic energy for the string endpoints, the Lagrangian describing this system is
$$
  L_{\perp}={1\over 2} \dot{x}_i^2 + {1\over 2} \dot{x}_j^2 -{1\over 2} (4g_{YM}^2 n_{ij}) (x_i-x_j)^2\, .
$$
The equations of motion (assuming the center of mass is at rest at the origin) are solved by
$$
  x_i =-x_j = A\sin (\sqrt{8g_{YM}^2 n_{ij}}t+\phi_0)\, .
$$
The energy of this solution is given by
$$
  E_{\perp}={1\over 2} \dot{x}_i^2 +{1\over 2} \dot{x}_j^2 +{1\over 2} (4g_{YM}^2 n_{ij}) (x_i-x_j)^2 = A^2 (8g_{YM}^2 n_{ij})
$$
which matches the anomalous dimensions.

\section{Strings between 3 giants} 

\begin{figure}[ht]
\begin{center}
\includegraphics[width=5.0cm]{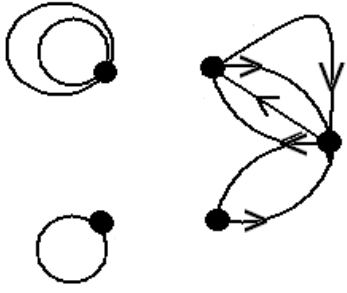}
\caption{The label $C$ for a system of 5 giants. $n_{12}=4$ strings stretch between branes 1 and 2 and $n_{23}=2$
strings stretching between branes 2 and 3. A string is attached to brane 4 and two strings are attached to 
brane 5.}
\label{fig:secgiantsys}
\end{center}
\end{figure}

In this section we consider the situation for which we have $n_{ij}$ strings stretching between giants $i$ and $j$, 
$n_{jk}$ strings stretching between giants $j$ and $k$ and $n_{ik}$ strings stretching between giants $i$ and $k$. 
See Figure \ref{fig:secgiantsys} for an example of the label $C$ when $p=5$ and $n_{ij}=4$, $n_{jk}=2$ and $n_{ik}=0$. 
The results of this section are also directly applicable to the case that any number of pairs and/or triples of 
mutually distinct branes have strings stretching between them. Just like in the last section, in this case the action
of the dilatation operator is given by a sum of terms which commute and can each be diagonalized using the same method.

\noindent
{\bf Construction of Creation and Annihilation Operators:}
In this case, to be general, we should introduce the parameters $n_{ij}$, $n_{ik}$ and $n_{jk}$
(repeated indices are not summed)
$$
  D = - g_{YM}^2 (n_{ij}\Delta_{ij}+n_{ik}\Delta_{ik}+n_{jk}\Delta_{jk})\, .
$$
For any label $C$, the Gauss Law implies that $n_{ij}+n_{ik}$ is even, $n_{ij}+n_{jk}$ is even and $n_{ik}+n_{jk}$ is even.
For a creation operator we again want (\ref{CO}). Make the ansatz
$$
  A^\dagger = aE_{ii}+bE_{ij}+cE_{ik}+dE_{ji}+eE_{jj}+fE_{jk}+gE_{ki}+hE_{kj}+iE_{kk}\, .
$$
Then (\ref{CO}) gives 3 $A^\dagger$s.
There is a nice analytic formula for the frequencies of these operators $\Omega_i =2g_{YM}^2\omega_i$ where
$$
  \omega_1 = 2\gamma,\qquad \omega_2 = n_{ik}+n_{ij}+n_{jk}+\gamma,\qquad \omega_3 = n_{ik}+n_{ij}+n_{jk}-\gamma,
$$
where
$$
  \gamma=\sqrt{n_{ij}^2+n_{ik}^2+n_{jk}^2-n_{ij}n_{ik}-n_{jk}n_{ik}-n_{ij}n_{jk}}\, .
$$
This proves that the spectrum of three giant system is indeed that of a set of oscillators.
For the frequency $\omega_1$ we find
$$
 A_1 = {\cal N}_1\Big[(n_{ij} - n_{ik})(n_{ik}-n_{ij}-\gamma)E_{ii}
$$
$$
  +\left( (n_{ik}-n_{ij})(n_{ik}-n_{ij}-\gamma) + (n_{ik}-n_{jk})(n_{ik}-n_{jk}-\gamma )
               \right)E_{ij}
$$
$$
    -(n_{ik}-\gamma -n_{jk})(n_{ik}-n_{jk})E_{ik}
    -(n_{ij}-n_{jk}+\gamma)(n_{ij}-n_{jk})E_{ji}
$$
$$
+(n_{ij}-n_{ik}+\gamma)(n_{ij}-n_{jk})E_{jj}
     +(n_{ij}-n_{jk})(n_{ik}-n_{jk})E_{jk}
$$
$$
+\left( (n_{jk}-n_{ij})(n_{jk}-n_{ij}-\gamma)+ (n_{ik}-n_{ij})(n_{ik}-n_{ij}-\gamma) \right)E_{ki}
$$
$$
    -( (n_{ij}-n_{ik})(n_{ij}-n_{ik}+2\gamma) +\gamma^2)E_{kj}
$$
$$
   -(n_{ij}-n_{ik}+\gamma)(n_{ik}-n_{jk})E_{kk}\Big]
$$
where
$$
  {\cal N}^{-2}_1 =(n_{ij}-n_{ik})^2 (n_{ik}-n_{ij}-\gamma)^2+((n_{ik}-n_{ij})(n_{ik}-n_{ij}-\gamma)+(n_{ik}-n_{jk})(n_{ik}-n_{jk}-\gamma))^2
$$
$$
  +(-n_{ik}+n_{jk}+\gamma)^2(n_{ik}-n_{jk})^2+(-n_{ij}+n_{jk}-\gamma)^2 (n_{ij}-n_{jk})^2+(n_{ij}-n_{ik}+\gamma)^2(n_{ij}-n_{jk})^2
$$
$$
+(n_{ij}-n_{jk})^2(n_{ik}-n_{jk})^2+((n_{jk}-n_{ij})(-n_{ij}+n_{jk}-\gamma)+(n_{ik}-n_{ij})(n_{ik}-n_{ij}-\gamma))^2
$$
$$
+(-(n_{ij}-n_{ik})(n_{ij}-n_{ik}+2\gamma)-\gamma^2)^2 +(n_{ik}-n_{ij}-\gamma)^2 (n_{ik}-n_{jk})^2\, .
$$
For the frequency $\omega_2$ we find
$$
 A_2 = {\cal N}_2 \Big( (n_{jk}-n_{ij}-\gamma)(E_{ii}+E_{ji}+E_{ki})+(n_{ij}-n_{ik}+\gamma)(E_{ij}+E_{jj}+E_{kj})
$$
$$
+(n_{ik}-n_{jk})(E_{ik}+E_{jk}+E_{kk})\Big)
$$ 
where
$$
  {\cal N}^{-1}_2 =\sqrt{6\gamma (2\gamma+2n_{ij}-n_{jk}-n_{ik})}\, .
$$
For the frequency $\omega_3$ we find
$$
A_3 = {\cal N}_3 \Big( (n_{jk}-n_{ij}+\gamma)(E_{ii}+E_{ji}+E_{ki})+
                               (n_{ij}-n_{ik}-\gamma)(E_{ij}+E_{jj}+E_{kj})
$$
$$
   +(n_{ik}-n_{jk})(E_{ik}+E_{jk}+E_{kk})\Big)
$$
where
$$
  {\cal N}_3^{-1}=\sqrt{6\gamma(2\gamma-2n_{ij}+n_{jk}+n_{ik})}\, .
$$
These oscillators close the following algebra
$$
  \big[ A_2,A_2^\dagger\big] ={4\over 3}\left(3N-3b_0-2b_1-b_2\right) +{1\over 3}
                              (\Delta_{ij}+\Delta_{ik}+\Delta_{jk})-P_2,\qquad \big[ A_2,A_3^\dagger\big] =-A_1
$$
$$
  \big[ A_2,A_1^\dagger\big] = A_3,\qquad \big[ A_2,A_3\big] =0=\big[ A_1,A_2 \big]
$$
$$
  \big[ A_3,A_3^\dagger\big] ={4\over 3}\left(3N-3b_0-2b_1-b_2\right) +
                              {1\over 3}(\Delta_{ij}+\Delta_{ik}+\Delta_{jk})-P_3\, ,\qquad \big[ A_3,A_1 \big] = A_2
$$
$$
  \big[ A_3,A_1^\dagger\big] = 0,\qquad  \big[ A_1,A_1^\dagger\big] = P_3-P_2\, ,
$$
where
$$
(4\gamma^2 - 2\gamma (n_{jk}+n_{ik}-2n_{ij})P_2 
= \left(n_{ij}-n_{jk}+\gamma \right) ^{2}E_{ii}
+ \left( n_{ij}-n_{ik}+\gamma \right) ^{2}E_{jj}
$$
$$
   +\left( n_{jk}-n_{ik} \right)  \left(n_{ij}-n_{jk}+\gamma \right)(E_{ik}+E_{ki})
   +\left( n_{jk}-n_{ij}-\gamma \right)  \left( n_{ij}-n_{ik}+\gamma \right)(E_{ji}+E_{ij})
$$
$$
   +\left( n_{ij}-n_{ik}+\gamma \right)  \left( n_{ik}-n_{jk} \right)(E_{jk}+E_{kj})
   +\left( n_{ik}-n_{jk} \right) ^{2}E_{kk} 
$$
and
$$
(4\gamma^2 + 2\gamma (n_{jk}+n_{ik}-2n_{ij})P_3 
= \left( n_{ij}-n_{jk}-\gamma \right) ^{2}E_{ii}
+\left( n_{ij}-n_{ik}-\gamma \right) ^{2}E_{jj}
$$
$$
+ \left( n_{ij}-n_{ik}-\gamma \right)  \left( n_{jk}-n_{ij}+\gamma \right)(E_{ij}+E_{ji})
+\left( n_{jk}-n_{ik} \right)  \left( -n_{jk}+n_{ij}-\gamma \right)(E_{ik}+E_{ki})
$$
$$
+ \left( n_{ik}-n_{jk} \right)  \left( n_{ij}-n_{ik}-\gamma \right)(E_{jk}+E_{kj})
+ \left( n_{ik}-n_{jk} \right) ^{2}E_{kk} \, .
$$
Note also that
$$
  \big[ A_2,A_2^\dagger\big] = \big[ A_3,A_3^\dagger\big] + \big[ A_1,A_1^\dagger\big]
$$
$$
\big[ P_2,A_2^\dagger\big]= A_2^\dagger,\qquad \big[ P_3,A_3^\dagger\big]= A_3^\dagger\, .
$$
Thus, if we set
$$
  A_1=\sqrt{3N-3b_0-2b_1-b_2}\sqrt{4\over 3}a_1,\qquad A_2=\sqrt{3N-3b_0-2b_1-b_2}\sqrt{4\over 3}a_2,
$$
$$
\qquad A_3=\sqrt{3N-3b_0-2b_1-b_2}\sqrt{4\over 3}a_3
$$
and consider the limit in which $\sqrt{3N-3b_0-2b_1-b_2}\sim\sqrt{N}\to\infty$ we find
$$
  \big[a_1,a_1^\dagger\big]=0,\qquad \big[a_2,a_2^\dagger\big]=1,\qquad \big[a_3,a_3^\dagger\big]=1
$$
and all other commutators vanish. Thus, we only have 2 oscillators. After keeping only the states that have a node
at $b_1=0$, we find that these oscillators have a frequency $4g_{YM}^2\omega_2$ and $4g_{YM}^2\omega_3$.

\noindent
{\bf Connection to Continuum Limit:}
We can again ask how this compares to the frequencies computed after we have taken the continuum limit of the $\Delta_{ij}$,
described in appendix H of \cite{Koch:2011hb}. From that appendix, we find
$$
 D=-g_{YM}^2 M_{ab}\left({\partial\over\partial x_a}{\partial\over\partial x_b}-{x_a x_b\over 4}\right)
$$
with
$$
M=\left[
\begin{array}{ccc}
n_{ij}+n_{ik} &-n_{ij}       &-n_{ik}\\
-n_{ij}       &n_{ij}+n_{jk} &-n_{jk}\\
-n_{ik}       &-n_{jk}       &n_{ik}+n_{jk}
\end{array}
\right]\, .
$$
The three frequencies are $\Lambda_i =2g_{YM}^2\lambda_i$, where
$$
  \lambda_1 = 0,\qquad \lambda_2 = n_{ik}+n_{ij}+n_{jk}+\gamma,\qquad \lambda_3 = n_{ik}+n_{ij}+n_{jk}-\gamma,
$$
and $\gamma$ is defined as above. The zero frequency again corresponds to the center of mass, which we fix.
Only the states with a node at $b_1=0$ will be retained, which doubles the above frequencies. 
Notice that the continuum limit has caught the full large $b_1$ spectrum.

\noindent
{\bf Classical Model:} Arguing exactly as we did in the last section leads to
$$
  L_{\perp}={1\over 2} \dot{x}_i^2 + {1\over 2} \dot{x}_j^2 + {1\over 2} \dot{x}_k^2  -{1\over 2} (2g_{YM}^2 n_{ij}) (x_i-x_j)^2
$$
$$
 -{1\over 2} (4g_{YM}^2 n_{ik}) (x_i-x_k)^2  -{1\over 2} (2g_{YM}^2 n_{jk}) (x_j-x_k)^2\, .
$$
The equations of motion are
$$
{d^2 x_i\over dt^2}=-2g_{YM}^2 n_{ij}(x_i -x_j)-2g_{YM}^2 n_{ik}(x_i -x_k)  \, ,
$$
$$
{d^2 x_j\over dt^2}= 2g_{YM}^2 n_{ij}(x_i -x_j)-2g_{YM}^2 n_{jk}(x_j -x_k)  \, ,
$$
$$
{d^2 x_k\over dt^2}= 2g_{YM}^2 n_{jk}(x_j -x_k)+2g_{YM}^2 n_{ik}(x_i -x_k)  \, .
$$
Again, fix the center of mass motion (the giant system is not moving anywhere).
It is easy to solve these equations; there are two normal modes. 
The energy of the solution with both modes excited, with amplitudes $A_1$ and $A_2$, is given by
$$
  E_{\perp}= A_1^2 8g_{YM}^2(n_{ik}+n_{ij}+n_{jk}+\gamma ) + A_2^2 8g_{YM}^2(n_{ik}+n_{ij}+n_{jk}-\gamma )
$$
which again matches the anomalous dimensions.

\section{Strings between 4 giants} 

The methods that we have outlined above work generally for any configuration $C$ of open strings. However, not surprisingly,
it becomes increasingly difficult to obtain simple analytic expressions. Obviously its a simple matter to get explicit numerical
results for any $C$. In this section we will simply write the equations one needs to obtain in the case that strings 
stretch in an arbitrary way between four giant gravitons.

\noindent
{\bf Construction of Creation and Annihilation Operators:} In this case, to be general, we should introduce 
the parameters $n_{ij}$, $n_{ik}$, $n_{il}$, $n_{jk}$, $n_{jl}$ and $n_{kl}$
$$
  D = -g_{YM}^2 (n_{ij}\Delta_{ij}+n_{ik}\Delta_{ik}+n_{il}\Delta_{il}+n_{jk}\Delta_{jk}+n_{jl}\Delta_{jl}+n_{kl}\Delta_{kl})
$$
For any $C$, $n_{ij}+n_{ik}+n_{il}$ is even, $n_{ij}+n_{jk}+n_{jl}$ is even,
$n_{ik}+n_{jk}+n_{kl}$ is even and $n_{il}+n_{jl}+n_{kl}$ is even. For a creation operator we again want (\ref{CO}).
This leads us to the eigenproblem of a 16$\times$16 matrix. For general parameters we get 6 $A^\dagger$s. Only three of
these survive in the large $b_{1,{\rm max}}$ limit. The frequencies of the oscillators which survive are roots of
$$
x^3-2(n_{ij}+n_{jl}+n_{ik}+n_{jk}+n_{kl}+n_{il})x^2
$$
$$
+(3n_{ik}n_{kl}+4n_{ij}n_{kl}+3n_{ij}n_{il}+3n_{ik}n_{il}+3n_{ik}n_{jk}+3n_{il}n_{jl}+3n_{kl}n_{il}+4n_{ik}n_{jl}+4n_{jk}n_{il}
$$
$$
+3n_{jk}n_{jl}+3n_{kl}n_{jl}+3n_{ij}n_{jk}+3n_{ij}n_{jl}+3n_{ij}n_{ik}+3n_{jk}n_{kl}) x
$$
$$
-4n_{ij}n_{kl}n_{jl}-4n_{ij}n_{jk}n_{kl}-4n_{ij}n_{ik}n_{kl}-4n_{ij}n_{kl}n_{il}-4n_{jk}n_{ik}n_{kl}
$$
$$
-4n_{jk}n_{ik}n_{il}-4n_{jk}n_{ik}n_{jl}-4n_{jl}n_{ik}n_{kl}-4n_{jl}n_{ik}n_{il}-4n_{jk}n_{kl}n_{il}-4n_{ij}n_{jk}n_{il}
$$
$$
-4n_{ij}n_{jk}n_{jl}-4n_{ij}n_{ik}n_{jl}-4n_{ij}n_{ik}n_{il}-4n_{jl}n_{jk}n_{il}-4n_{jl}n_{kl}n_{il}=0\, .
$$
It is now straight forward to construct the algebra of the resulting oscillators as well as their large $b_1$ limit.
We again find that this result is consistent with both the continuum limit of $D$ (as outlined in appendix H of \cite{Koch:2011hb})
and the classical model of masses and springs. This computation (as well as the extension to situations in which strings interconnect
more than 4 giants) is straight forward but a little tedious. 

In summary, two things have been achieved in this note.
The continuum limit of the dilatation operator was obtained in appendix H of \cite{Koch:2011hb}. What is the relation
between the study of \cite{Koch:2011hb} and our result here? In \cite{Koch:2011hb} the large $b_{1,{\rm max}}$ limit was taken and the resulting eigenvalue problem was solved. Here we have first solved the eigenvalue problem and have then taken the large $b_{1,{\rm max}}$ limit.
Our result is in perfect agreement with the continuum limit obtained in \cite{Koch:2011hb}, and justifies the use of the simple Harmonic oscillator Hamiltonian obtained there. In particular, in the continuum limit the variables $s_i$ become continuous coordinates and the operators of a good scaling dimension are obtained by summing restricted
 Schur polynomials with coefficients given by the harmonic oscillator wave functions.
The second thing we have achieved is that the values of the anomalous dimensions have been reproduced by the normal mode frequencies
of a coupled system of open strings. This provides non-trivial support for their interpretation in the dual theory as excited
giant gravitons.

\noindent
{\it Acknowledgements:}
We would like to thank Kevin Goldstein for pleasant discussions. 
This work is based upon research supported by the South African Research Chairs
Initiative of the Department of Science and Technology and National Research Foundation.
Any opinion, findings and conclusions or recommendations expressed in this material
are those of the authors and therefore the NRF and DST do not accept any liability
with regard thereto.

\end{document}